\documentclass[journal,draftclsnofoot,onecolumn,12pt]{IEEEtran}

\usepackage{amsthm,amssymb,amsmath,graphicx,multirow,color,amsfonts}
\usepackage[update,prepend]{epstopdf}
\usepackage{amsmath,amsfonts}
\usepackage{algorithmic}
\usepackage{algorithm}
\usepackage{array}
\usepackage{textcomp}
\usepackage{stfloats}
\usepackage{url}
\usepackage{verbatim}
\usepackage{graphicx}
\usepackage{cite}
\usepackage[latin1]{inputenc}
\usepackage{tikz}
\usepackage{bbm} 
\usepackage{pdfpages}
\usepackage{multirow}
\usepackage{graphicx}
\usepackage{multicol}
\usepackage[justification=centering]{caption}
\usepackage{textcomp}
\usepackage{psfrag}
\usepackage{arydshln}
\usepackage{url}
\usepackage{soul}
\usepackage{graphicx,color}
\usepackage{comment}
\usepackage{subfigure}
\usepackage[nolist]{acronym}
\usepackage{algorithm,algorithmic} 
\usepackage{mathtools,lipsum}
\usepackage{cuted}
\usepackage{amsmath}
\usepackage{mathrsfs}
\usepackage{color}
\usepackage{txfonts}
\usepackage{float}
\usepackage{balance} 
\usepackage[shortlabels]{enumitem}
\usepackage[capitalise]{cleveref}
\Crefname{equation}{Eq.\!}{Eqs.\!}
\Crefname{figure}{Fig.\!}{Figs.\!}
\Crefname{tabular}{Tab.\!}{Tabs.\!}
\Crefname{section}{Section\!}{Sections.\!}

\allowdisplaybreaks 

\usepackage{setspace}	

\begin{document}

\newtheorem{lemma}{Lemma}
\newtheorem{thm}{Theorem}
\newtheorem{definition}{Definition}
\newtheorem{ndef}{Definition}
\newtheorem{nrem}{Remark}
\newtheorem{theorem}{Theorem}
\newtheorem{corollary}{Corollary}
\newtheorem{example}{Example}
\newtheorem{remark}{Remark}
\newtheorem{assumption}{Assumption}
\newtheorem{approximation}{Approximation}
\newtheorem{proposition}{Proposition} 

\graphicspath{{./Figures/}}
	\begin{acronym}

\acro{5G-NR}{5G New Radio}
\acro{3GPP}{3rd Generation Partnership Project}

\acrodef{awgn}[AWGN]{Additive White Gaussian Noise}     
\acro{AWGN}{additive white Gaussian noise}
\acro{AN}{Artificial Noise}
\acro{ES}{eavesdropper satellite}
\acro{LS}{legitimate satellite}
\acro{UAV}{unmanned aerial vehicles}
\acro{BEP}{bit error probability}
\acro{BER}{bit error rate}
\acro{BS}{base station}
\acro{BW}{BandWidth}

\acrodef{cdf}[CDF]{cumulative distribution function}   
\acro{CDF}{Cumulative Distribution Function}
\acrodef{c.d.f.}[CDF]{cumulative distribution function}
\acro{CCDF}{complementary cumulative distribution function}
\acrodef{ccdf}[CCDF]{complementary CDF}               
\acrodef{c.c.d.f.}[CCDF]{complementary cumulative distribution function}

\acro{CDMA}{Code Division Multiple Access}
\acro{ch.f.}{characteristic function}
\acro{CSI}{channel state information}

\acro{PLS}{Physical Layer Security}

\acro{dft}[DFT]{discrete Fourier transform}

\acro{DL}{DownLink}

\acro{DFT-s-OFDM}{Discrete Fourier Transform-spread-Orthogonal Frequency Division Multiplexing}

\acro{FFT}{fast Fourier transform}
\acro{FDMA}{Frequency Division Multiple Access}

\acro{GA}{Gaussian approximation}
\acro{GW}{Gateway}

\acro{HAP}{high altitude platform}
\acro{HetNet}{Heterogeneous network}

\acro{IFFT}{inverse fast Fourier transform}
\acro{i.i.d.}{independent, identically distributed}
\acro{IoT}{Internet of Things}                      
\acro{IR}{impulse radio}

\acro{IEEE}{Institute of Electrical and Electronics Engineers}

\acro{LEO}{Low Earth Orbit}
\acro{LF}{likelihood function}

\acro{LoS}{Line-of-Sight}
\acro{LRT}{likelihood ratio test}

\acro{LPWAN}{Low Power Wide Area Network}
\acro{LoRaWAN}{Low power long Range Wide Area Network}
\acro{NLoS}{Non-Line-of-Sight}
\acro{GRV}{Gamma Random Variable}
\acro{MGF}{Moment Generating Function}
\acro{NB}{narrowband}
\acro{NBI}{narrowband interference}
\acro{NLA}{nonlinear sparse approximation}
\acro{NLOS}{Non-Line of Sight}

\acro{LOS}{Line of Sight}

\acro{OFDMA}{Orthogonal Frequency-division Multiple Access}

\acro{PDF}{Probability Density Function}
\acrodef{p.d.f.}[PDF]{probability distribution function}

\acro{PMF}{probability mass function}                             
\acrodef{p.m.f.}[PMF]{probability mass function}
\acro{PP}{Point Process}
\acro{PPP}{Poisson Point Process}
\acro{BPP}{Binomial Point Process}
\acro{PCP}{Poisson Cluster Process}
 \acro{PGFL}{Probability Generating Functional}
\acro{MCP}{Matern Cluster Process}

\acro{LT}{Laplace Transform}

\acro{RV}{Random Variable}

\acro{RB}{resource block}

\acro{SINR}{signal-to-interference and noise ratio}
\acro{SIR}{signal-to-interference ratio}

\acro{SNR}[\textrm{SNR}]{signal-to-noise ratio} 

\acro{SR}{Shadowed-Rician}

\acro{SG}{Stochastic Geometry}

\acro{TBS}{terrestrial base station}


\acro{PL}{path-loss}

\acro{UE}{User Equipment}
\acro{UL}{UpLink}

\acro{WLAN}{wireless local area network}
\acro{wm}[WM]{Wishart matrix}                               
\acroplural{wm}[WM]{Wishart matrices}
\acro{WMAN}{wireless metropolitan area network}
\acro{WPAN}{wireless personal area network}
\acro{WSN}{wireless sensor network}                        
\acro{WSS}{Wide-Sense Stationary}
\acro{WHO}{World Health Organization}
\acro{Wi-Fi}{Wireless Fidelity}

\acro{VLC}{Visible Light Communication}
\acro{VPN}{Virtual Private Network} 
\acro{RF}{Radio Frequency}
\acro{FSO}{Free Space Optics}
\acro{IoST}{Internet of Space Things}

\acro{GSM}{Global System for Mobile Communications}
\acro{2G}{Second-generation cellular network}
\acro{3G}{Third-generation cellular network}
\acro{4G}{Fourth-generation cellular network}
\acro{5G}{Fifth-generation cellular network}	
\acro{gNB}{next-generation Node-B Base Station}
\acro{NR}{New Radio}
\acro{UMTS}{Universal Mobile Telecommunications Service}
\acro{LTE}{Long Term Evolution}

\acro{QoS}{Quality of Service}
\end{acronym}
	
\newcommand{\SAR} {\mathrm{SAR}}
\newcommand{\WBSAR} {\mathrm{SAR}_{\mathsf{WB}}}
\newcommand{\gSAR} {\mathrm{SAR}_{10\si{\gram}}}
\newcommand{\Sab} {S_{\mathsf{ab}}}
\newcommand{\Eavg} {E_{\mathsf{avg}}}
\newcommand{\ft}{f_{\textsf{th}}}
\newcommand{\alphatf}{\alpha_{24}}

\title{Enhancing Physical Layer Security in \\ LEO Satellite-Enabled IoT Network Communications}
\author{
Anna Talgat, {\em Student Member, IEEE}, Ruibo Wang, \\ Mustafa A. Kishk, {\em Member, IEEE}, and Mohamed-Slim Alouini, {\em Fellow, IEEE}
\thanks{Anna Talgat, Ruibo Wang, and Mohamed-Slim Alouini are with KAUST, CEMSE division, Thuwal 23955-6900, Saudi Arabia. Mustafa Kishk is with the Department of Electronic Engineering, Maynooth University, W23 F2H6, Ireland. (e-mail:  anna.talgat@kaust.edu.sa; ruibo.wang@kaust.edu.sa; mustafa.kishk@mu.ie; slim.alouini@kaust.edu.sa. Corresponding author: Ruibo Wang.)}
\vspace{-4mm} }
\maketitle

\begin{abstract}
The extensive deployment of Low Earth Orbit (LEO) satellites introduces significant security challenges for communication security issues in  Internet of Things (IoT) networks. With the rising number of satellites potentially acting as eavesdroppers, integrating Physical Layer Security (PLS) into satellite communications has become increasingly critical.
However, these studies are facing challenges such as dealing with dynamic topology difficulties, limitations in interference analysis, and the high complexity of performance evaluation. To address these challenges, for the first time, we investigate PLS strategies in satellite communications using the Stochastic Geometry (SG) analytical framework. We consider the uplink communication scenario in an LEO-enabled IoT network, where multi-tier satellites from different operators respectively serve as legitimate receivers and eavesdroppers. In this scenario, we derive low-complexity analytical expressions for the security performance metrics, namely availability probability, successful communication probability, and secure communication probability. By introducing the power allocation parameters, we incorporate the Artificial Noise (AN) technique, which is an important PLS strategy, into this analytical framework and evaluate the gains it brings to secure transmission. In addition to the AN technique, we also analyze the impact of constellation configuration, physical layer parameters, and network layer parameters on the aforementioned metrics.
\end{abstract}
 

\section{Introduction}
\subsection{Motivation}
\ac{LEO} satellites play a crucial role in enhancing the connectivity of the \ac{IoT} devices, especially in remote and underserved areas where traditional terrestrial networks struggle to provide coverage~\cite{9210567, 8002583, 9442378}. The rapid expansion of satellite launches, which has been growing by 30\% annually since 2012, is due to advancements in rocket launch platforms~\cite{8976900}. This brings enormous opportunities for the connectivity of \ac{IoT} devices but also poses new challenges for communication security. 
\par The deployment of mega-constellations makes more satellites act as potential eavesdropping threats, especially for uplink transmission from \ac{IoT} devices to satellites.  Furthermore, due to the dynamic nature and high speed of \ac{LEO} satellite constellations relative to the Earth, there are frequent changes in the signal quality received by both the primary and potential eavesdropping channels. Atmospheric conditions, multipath fading, and interference further impact the signal. At certain times, the main channel may experience poor conditions, resulting in a weaker received signal. Simultaneously, an eavesdropper may be in a location with better channel conditions, leading to it experiencing a stronger signal. For example, rain fade can weaken the main channel signal, while an eavesdropper in a clear sky region may experience less attenuation and, therefore, better signal quality. This variability can lead to scenarios where the eavesdropper receives a higher quality signal than the main communication link, thus posing a significant security threat.
Therefore, the urgent need for enhancing uplink security strategies should be emphasized. 
\par To address these challenges, our research integrates the \ac{SG} analytical framework with advanced \ac{PLS} techniques. This novel combination enables comprehensive analysis of dynamic topology and interference, providing low-complexity solutions for evaluating security performance in \ac{IoT} networks based on multi-tier \ac{LEO} satellite constellations.

\subsection{PLS Techniques in Satellite Networks}
Recently, \ac{PLS} techniques have been increasingly recognized as potential solutions for enhancing security within satellite networks. The idea of \ac{PLS} originates from Shannon's foundational concept about perfect secrecy.  Subsequently, authors in \cite{wyner1975wire} proposed the eavesdropper channel model, considering that the information received by the eavesdropper is a degraded version of the legitimate received information.
\par The core idea of applying \ac{PLS} in satellite networks is to use the randomness of wireless channels, such as interference, fading, and noise, to achieve secure transmission \cite{8850067, tedeschi2022satellite,9781300,abdelsalam2023physical}. A notable example is the \ac{AN} technique, which was first proposed in \cite{4543070}. 
The principle of \ac{AN} technology is that the \ac{IoT} device generates an additional interference signal while transmitting useful signals \cite{7906491}. This interference signal can be filtered out by legitimate receiving satellites during demodulation, thereby affecting only the eavesdropping satellites \cite{6108297}. As a result, \ac{IoT} devices reduce the signal quality of the eavesdropping link by sacrificing some of their own power to transmit \ac{AN}.
\par
Several pioneering studies have advanced the integration of \ac{PLS} techniques within satellite communications. Authors in \cite{8300047} provided foundational frameworks by optimizing resources for secure satellite communications across different receiver scenarios, including fixed and mobile units, while \cite{8571220} provided a detailed theoretical analysis on securing non-geostationary orbit satellite downlinks. Further, research like \cite{9028250} introduced threshold-based user scheduling to improve secrecy in multiuser systems, and \cite{10194586} proposed cooperative secrecy tactics in LEO-integrated networks through jamming signals and optimized power allocation. Moreover, \cite{9463880} explored the role of \ac{UAV}s in strengthening satellite-vehicular communication security, and \cite{8902531} leveraged time-packing techniques to safeguard \ac{IoT} devices' direct satellite links.
Additionally, recent advancements in \ac{PLS} techniques include secrecy-energy efficient hybrid beamforming for satellite-terrestrial integrated networks to improve security and energy efficiency \cite{9453840}, signal-to-leakage-plus-noise ratio (SLNR)-based secure energy-efficient beamforming applied in multibeam satellite systems \cite{9827562}, and refracting reconfigurable intelligent surface (RIS)-aided hybrid satellite-terrestrial relay networks developed for joint beamforming design and optimization \cite{9726800}. Furthermore, rate-splitting multiple access, which indirectly supports \ac{PLS} by improving interference management and signal reliability, has been used to support \ac{IoT} in satellite and aerial-integrated networks \cite{9324793}. These studies demonstrate the wide range of \ac{PLS} applications in safeguarding satellite communications against potential security threats, highlighting the most recent developments in improving the security and efficiency of satellite and \ac{IoT} networks.

\par  Although the issue of \ac{PLS} has gained wide attention in the field of satellite communications, there are still some challenges that remain unresolved. The first challenge is that the satellite network's topology is dynamic; thus, the number of eavesdropping and legitimate satellites follow a specific stochastic distribution. Furthermore, the spatial distribution of satellites also greatly influences the probability of eavesdropping or successful transmission \cite{jung2022performance}.  The second challenge is that the deployment of a large number of devices makes the interference hard to ignore \cite{9313025}. Finally, as the number of satellites and \ac{IoT} devices increases, evaluating the effectiveness of \ac{PLS} strategies becomes increasingly challenging. This is because more device locations need to be generated, and more \ac{IoT}-satellite links need to be simulated \cite{8300047,10194586}. Therefore, the third challenge lies in evaluating the performance of \ac{PLS} strategies with low computational complexity. 

\subsection{Related Works for Stochastic Geometry}
\ac{SG} is a mathematical tool that can perfectly address the above challenges. It is one of the most suitable methods for analyzing large-scale random network topologies \cite{lou2023coverage}, and the \ac{SG}-based analytical framework can account for stochastic variations in interference~\cite{haenggi2012stochastic}. The analytical expressions derived within the \ac{SG} framework can be represented as functions of system parameters such as the number of satellites \cite{wang2023reliability}. Therefore, we can achieve a low-complexity mapping from system parameters to performance metrics, and the complexity does not increase with the scale of the constellation \cite{9684552}. Next, we will introduce the research related to satellites and \ac{PLS} in the field of \ac{SG}.

\par
Modeling satellites through the \ac{SG} model is an emerging research method. \ac{BPP} is one of the most widely used satellite modeling methods \cite{9177073,lou2023haps}. The topological performance (such as the distribution of distances between communication parties \cite{9841569}) and the network-level performance (such as coverage probability and data rate \cite{okati2020downlink,wang2024ultra}), derived from \ac{BPP}, align well with actual satellite constellation outcomes. Compared to \ac{BPP}, \ac{PPP} is more suitable for modeling networks in non-closed areas \cite{9755277,10463093}, such as ground-based \ac{IoT} networks. Among \ac{SG}-based satellite works, studies most closely related to our paper are~\cite{kim2023ergodic,jung2022satellites} and our previous work \cite{9218989}. Our past research \cite{9218989} involved addressing coverage issues in multi-tier satellite networks, where contact distance and association strategies laid the groundwork for this paper. However, \cite{9218989} did not involve \ac{IoT} devices or \ac{PLS} issues.  While \cite{kim2023ergodic} studied \ac{PLS} in the downlink \ac{LEO} satellite networks, focusing on the spatial distribution of satellites, users, and potential eavesdroppers modeled using homogeneous spherical \ac{PPP} with eavesdroppers located on the ground. 
As for \cite{jung2022satellites}, authors explored the effect of single-tier satellite eavesdroppers on uplink transmission. The study provided support and inspiration for the research scenario of this paper. Unlike our paper, however, authors in~\cite{jung2022satellites} did not consider \ac{PLS} techniques, nor did it involve modeling of legitimate constellations. Therefore, our study significantly extends the scope beyond \cite{jung2022satellites}.

\par
Additionally, some literature has utilized the \ac{SG} framework to analyze the secure communication performance of planar networks \cite{7990366,9458036}. Authors in \cite{6512533} initiated \ac{PLS} methodology within cellular networks, and authors in \cite{9627722} further explored this concept within millimeter-wave relaying networks. Authors in \cite{9458036} expanded the \ac{SG} framework to \ac{IoT} networks, where sensors, access points, and eavesdroppers are modeled by  \ac{PPP}s. Furthermore, authors in \cite{7906491} and \cite{7839915} introduced \ac{AN} technique into secure transmission in terrestrial networks, highlighting the significance of utilizing spatial randomness and strategic \ac{AN}  allocation. 
Although these studies have laid a solid foundation for researching \ac{PLS} strategies in the \ac{SG} field within ground networks, investigating \ac{PLS} issues in satellite networks still faces many challenges. 
The spherical topology, channel models, and performance metrics, such as availability in satellite networks, are significantly different from those in ground networks.

\subsection{Contribution}
As the first study for \ac{PLS} in satellite networks under  \ac{SG}, the specific contributions are as follows:

\begin{itemize}
     \item We introduce a novel multi-tier satellite network topology for uplink transmission, where one tier consists of legitimate satellites while other tiers consist of potential eavesdroppers. This topology, which includes satellites at various altitudes serving distinct roles, accurately reflects the complexity of real-world \ac{LEO} satellite constellations.
    \item We develop an analytical framework using \ac{SG} to model the spatial distribution and interaction of satellites and \ac{IoT} devices. This framework allows us to derive low-complexity analytical expressions for key performance metrics: satellite availability probability, coverage probability, secrecy outage probability, successful communication probability, and secure communication probability.
    \item By incorporating the \ac{AN} technique within our SG-based analytical framework, we enhance physical layer security. This technique, with the strategic allocation of power to  \ac{AN}, degrades the signal quality perceived by potential eavesdroppers while maintaining the communication quality for legitimate receivers, thereby improving secure transmission performance.
    \item We conduct Monte Carlo simulations to validate the accuracy of our analytical results. These simulations confirm that the derived expressions can serve as accurate and low-complexity alternatives to more computationally intensive simulations for performance evaluation.
    \item We reveal useful system-level insights regarding the impact of constellation configuration (such as the number of satellites and their altitude), physical layer parameters (such as beamwidth), and network layer parameters (such as \ac{IoT} device density) on availability probability and secure communication probability. These insights offer practical guidelines for optimizing secure \ac{LEO} satellite-enabled \ac{IoT} networks.
\end{itemize}

\begin{figure*}[!ht]
\centering
\includegraphics[width=\textwidth]{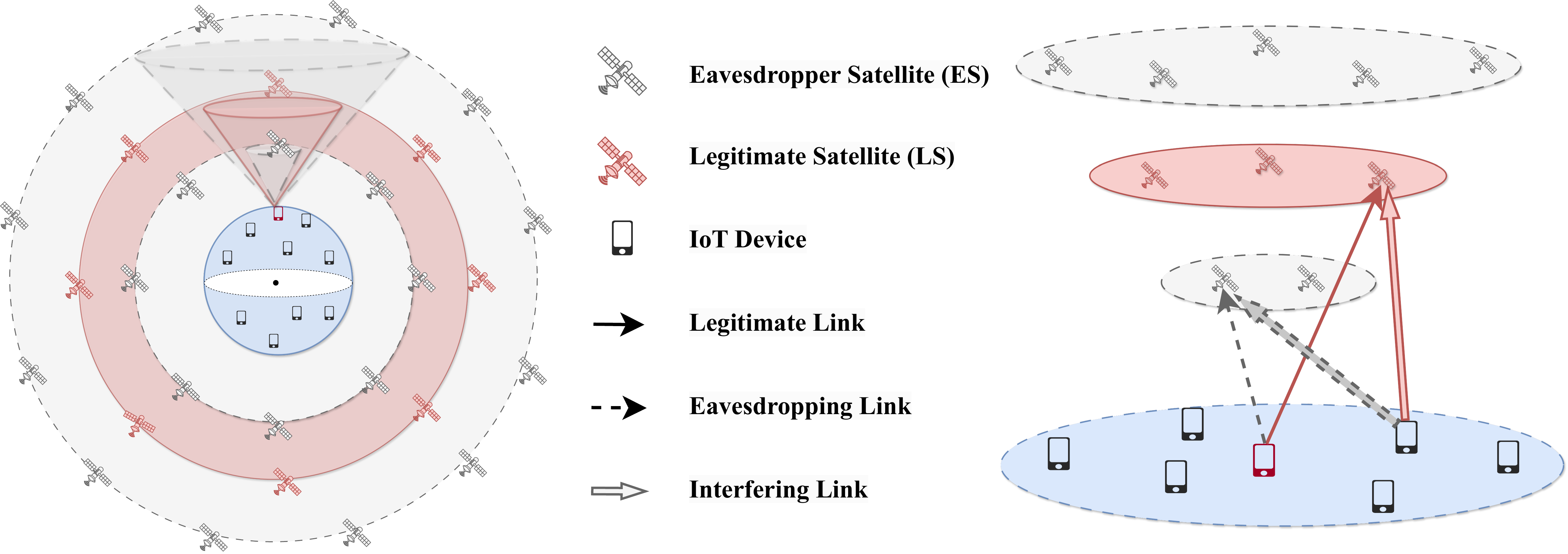}
\caption{Illustration of the multi-tier architecture of the network, depicting LS and ES at various altitudes. This configuration enables detailed analysis of signal interactions and security measures across d different altitude levels.} 
\label{system figure}
\end{figure*}

\section{System Model}
This section presents the system model for our investigation into enhancing the security of \ac{IoT} communications through \ac{LEO} satellites using \ac{PLS} techniques. We consider uplink transmission in a multi-tier \ac{LEO} satellite network with terrestrial \ac{IoT} devices. In addition, we employ \ac{AN} as a countermeasure against potential eavesdropping threats. The symbols, key abbreviations, and their meanings are listed in TABLE~\ref{tab:notations_abbreviations}. 

\subsection{Spatial Distribution Model} 
The \ac{IoT} devices are distributed across the Earth's surface, following a homogeneous \ac{PPP} model, which closely reflects real-world conditions. The \ac{PPP} is denoted as $\Phi_{u}\equiv\{\mathbf{u}_i\} \in \mathbb{R}^3$ with density $\lambda_{u}$. \ac{IoT} devices transmit their data to the designated \ac{LEO} satellites. Without loss of generality, the typical \ac{IoT} device is assumed to be located at $\mathbf{u_0}=\left( 0, 0, R_\oplus \right)$ in the spherical system, where the $R_\oplus$ is the radius of the Earth ($\approx 6371$ km). Apart from the typical device, other \ac{IoT} devices act as potential sources of interference in the uplink transmission.

\par
We consider a $K$-tier \ac{LEO} satellite constellation with different altitudes $a_k$, where $k=1, \dots, K$. Satellites from the $k$-th tier follow a homogeneous \ac{BPP} on the surface of a sphere $S_k$ with radius $R_k=a_k+R_\oplus$, where the positions of $N_k$ satellites in the $k$-th tier are \ac{i.i.d.}. The sphere $S_k$ can be expressed with spherical coordinates as $S_k=\{R_k, 0\leq \theta \leq \pi, 0\leq \varphi \leq 2 \pi\}$, where $\theta$ and $\varphi$ are polar angle and azimuthal angle, respectively. Finally, we denote the location of satellites on $k$-th sphere, as $\Phi_{S_k}\equiv\{\mathbf{x}_{kj}\}$ where  $\mathbf{x}_{kj}\in \mathbb{R}^3$ indicates the location of $j$-th satellite on $k$-th tier. As shown in Fig.~\ref{system figure}, the satellites in the constellation serving the \ac{IoT} devices are \ac{LS}s, while other satellites potentially intercepting communications from \ac{IoT} devices are known as  \ac{ES}s. The $K$-tier \ac{LEO} satellite constellation includes a legitimate tier composed of \ac{LS}s, as well as $K-1$ tiers consisting of \ac{ES}s. 
\par The multi-tier architecture, as illustrated in Fig.~\ref{system figure}, represents a realistic deployment scenario where satellites are positioned at varying altitudes. This structure allows for an in-depth analysis of signal propagation and interception across different layers, enhancing the practical relevance of our study. By considering satellites at multiple altitudes, we can evaluate complex interaction dynamics and develop more effective \ac{PLS} strategies.

\begin{table}[t!]
\caption{Notations and Abbreviations}
\label{tab:notations_abbreviations}
\centering
\resizebox{0.65\textwidth}{!}{
\begin{tabular}{c|c} 

\hline
\textbf{Symbol/Abbr.} & \textbf{Description} \\
\hline
\hline LEO & Low Earth Orbit \\
\hline PLS & Physical Layer Security \\
\hline SG & Stochastic Geometry \\
\hline AN & Artificial Noise \\
\hline BPP & Binomial Point Process \\
\hline PPP & Poisson Point Process \\
\hline LS & Legitimate Satellite \\
\hline ES & Eavesdropper Satellite \\
\hline $\Phi_{S_k}\equiv\{\mathbf{x}_{kj}\}  $  & \ac{BPP} for satellites in $k$-th tier \\
\hline $\Phi_u \equiv \{\mathbf{u}_{i}\} $ & \ac{PPP} for \ac{IoT} devices locations \\
\hline \(\mathcal{A}_{m, \rm vis}\) & Visibility area for the $m$-th tier  \\ 
\hline $\mathcal{I}_m $ & Interference power for the serving \ac{LS} at  tier  $m^{\rm}$ \\
\hline $\mathcal{I}_{kj}$ & Interference power for $j$-th \ac{ES} at  tier  $k$ \\
\hline $\mathcal{L}_{\mathcal{I}}(\cdot)$ &  Laplace transform of the interference $\mathcal{I}$ \\
\hline $\mathcal{P}_{m, \rm av}$ & Availability probability for $m$-th tier\\
\hline $\mathcal{P}_{m, \rm cov}$& Coverage probability for $m$-th tier \\
\hline $\mathcal{P}_{m, \rm suc}$  & Successful communication probability for $m$-th tier \\
\hline$\mathcal{P}_{\rm out}$ & Secrecy outage probability \\
\hline $\mathcal{P}_{\rm sec} $ &  Secure communication probability \\
\hline
\hline
\end{tabular}
}
\end{table}

\subsection{Channel Model}
\begin{figure}[!t]
\centering
\includegraphics[width=0.7\textwidth]{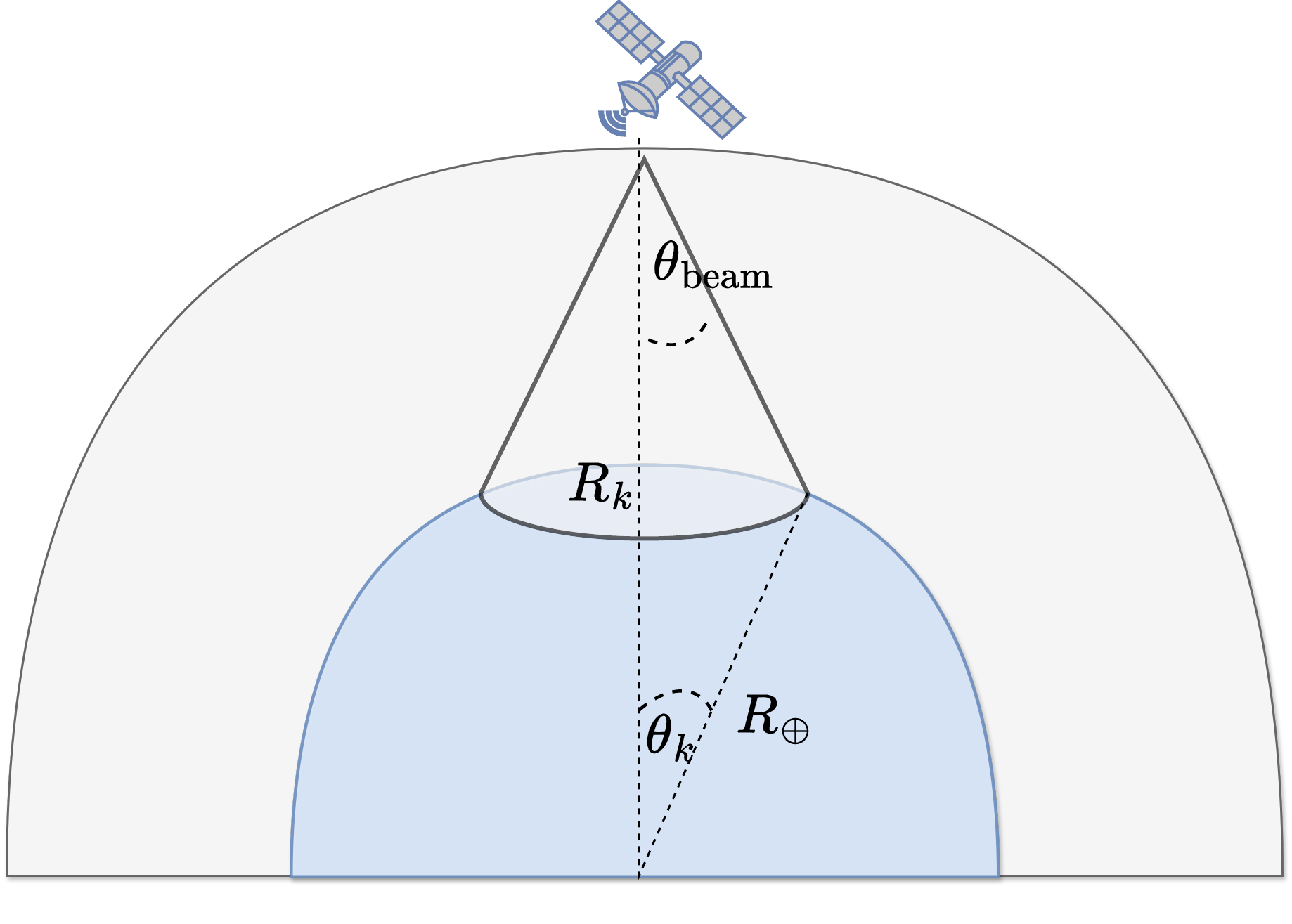}
\caption{Illustration of satellite beamwidth \(\theta_{\rm beam}\) and its impact on Earth's surface coverage.}
\label{beamwidth angle figure}
\end{figure}

In our channel model, we combine the effects of large-scale and small-scale fading to represent signal propagation accurately. The large-scale fading component, modeled by the free-space path loss, accounts for the overall signal attenuation due to the distance between the transmitter and the receiver. This component is expressed as $\left(\frac{c}{4 \pi f_c d_{kj, i}} \right)^2$, where  $f_c$ indicates the carrier frequency, $c$ is the speed of light and $d_{kj, i}$ is the distance between the $i$-th \ac{IoT} device and the $j$-th satellite in the $k$-th tier, 
\begin{align}\label{distance eqn}
d_{kj,i}=\sqrt{R_\oplus^2+R_k^2-2R_\oplus R_k \cos{\theta_{kj,i}}},   
\end{align}
where $\theta_{kj,i}$ is the central angle between the $i$-th \ac{IoT} device and the $j$-th satellite. An example of the central angle is shown in Fig.~\ref{beamwidth angle figure}.
\par We adopt the Shadowed-Rician distribution to model small-scale fading, as it provides a realistic and precise approximation of satellite-terrestrial channels, accurately capturing small-scale variations and reflecting real-world conditions~\cite{loo1985statistical, abdi2003new,jung2022performance}. Thus, the small-scale fading component, represented by $|h_{kj,i}|^2$, follows the Shadowed-Rician distribution, which describes the fading between the $i$-th \ac{IoT} device and the $j$-th satellite in the $k$-th tier. 
\par Combining these components, the received signal power $P_{\rm rec}$, at the satellite during uplink transmission is modeled as 
\begin{align}\label{P_rec}
P_{\rm rec}\left(d_{kj, i}\right)=  P_t \left(\frac{c}{4 \pi f_c d_{kj, i}} \right)^2 G |h_{kj, i}|^2, 
\end{align}
where $P_t$ denotes the transmit power of the \ac{IoT} device, $G$ is the product of the gains of the transmitting and receiving antennas.
\par Next, we determine the range for the central angle $\theta$. Firstly, we consider that \ac{IoT} devices located within the satellite receiving beam's main lobe can establish a stable link with the satellite. The satellite beam is pointed towards the center of the Earth, and the main lobe angle of the receiving beam is denoted as $2 \theta_{\rm beam}$. As illustrated in Fig.~\ref{beamwidth angle figure}, $\theta_{\rm beam}$ is the half beamwidth angle of the satellite. In addition, assuming that only satellites with line-of-sight links established with \ac{IoT} devices can communicate, the upper bound of the central angle can be represented as,
\begin{align}\label{eq: theta_max}
    \theta_{k, \rm max}=
    \begin{dcases}
    \sin^{-1}{\left(\frac{R_k}{R_\oplus}\sin{\theta_{\rm beam}}\right)}-\theta_{\rm beam}, & \theta_{\rm beam} <\sin^{-1}{\left(\frac{R_\oplus}{R_k}\right)}, \\
 \cos^{-1}{\left(\frac{R_\oplus}{R_k}\right)}, & \theta_{\rm beam} \geq \sin^{-1}{\left(\frac{R_\oplus}{R_k}\right)},
    \end{dcases}
\end{align}
and the range for $\theta$ is $0< \theta < \theta_{k, \rm max}$.
\par As mentioned earlier, the Shadowed-Rician distribution is used to model small-scale fading. However, due to the complexity of deriving closed-form solutions from the Shadowed-Rician channel model, we adopt the Gamma fading model as a practical approximation for the Shadowed-Rician fading~\cite{10463093}. Thus, the \ac{CDF} of small-scale fading $|h|^2$ is tightly bounded by as follows
 \begin{equation} \label{Upper Gamma} 
 F_{|h|^2}(x)\leq \left[1-\exp\left({-\frac{(m_1!)^{-\frac{1}{m_1}}}{m_2}x}\right)\right]^{m_1} \quad \text{if} \quad m_1 \leq 1, \\
\end{equation}
where $m_1$ and $m_2$ denote shape and scale parameters, respectively. 

\subsection{Artificial Noise}
In this article, we introduce \ac{AN}  into the transmitted signal to enhance its secrecy. The transmit power of \ac{IoT} device, $P_t$, is divided into two parts: a fraction $\gamma P_t $ is allocated for transmitting confidential information, $\left(1-\gamma\right) P_t $ is dedicated to transmitting \ac{AN}. $\gamma$ is known as the information-bearing ratio, which falls within the interval $[0,1]$. In scenarios where $\gamma=1$, \ac{AN}  is not applied, and all of the \ac{IoT} device's transmit power is dedicated to signal transmission. 
\par Then, we can present the communication links into the legitimate link (typical \ac{IoT} device to \ac{LS}), eavesdropping link (typical \ac{IoT} device to \ac{ES}), and interfering link (interfering \ac{IoT} device to the satellite of interest). Due to \ac{LS}s knowing in advance that typical \ac{IoT} devices employ \ac{AN}  strategies, the \ac{AN}  can be filtered out by \ac{LS}s. When the \ac{IoT} device is served by the $m$-th tier, the \ac{SINR} attained at the serving satellite is expressed as
\begin{align}\label{sinr for LS}
    {\rm SINR}_{\rm LS}=\frac{\gamma  P_{\rm rec}\left(d_{m0, 0}\right)}{\mathcal{I}_m+\sigma_{\rm S}^2}, 
\end{align}
where \(P_{\rm rec}(d_{m0, i})\) represents the received power at the serving satellite from both typical (\(i=0\)) and interfering (\(i\neq 0\)) devices, with the distance function $d_{m0, i}$ defined in (\ref{distance eqn}). For notation simplicity, and because all \ac{IoT} devices (typical or interfering) communicate only with the serving satellite in the legitimate tier $m$, the subscript $0$ used for the serving satellite is omitted from $d_{m0,i}$ in favor of  $d_{m,i}$, $\forall i$.
The noise power at the satellite is denoted by \(\sigma_{\rm S}^2=N_0 \times B\), where $N_0$ is the noise spectral density and $B$ is the bandwidth. Furthermore, 
\begin{equation}
\mathcal{I}_m=\sum\limits_{i\in \Phi_u\setminus \{i=0\}}  P_{\rm rec}(d_{m, i})
\end{equation}
denotes the cumulative interference from other \ac{IoT} devices (\(i\neq 0\)) at the serving satellite.
\par \ac{ES}s, in contrast, are incapable of filtering out the \ac{AN}. For the $j$-th \ac{ES} located in the $k$-th tier (\(\forall k \neq m\)), the \ac{SINR} of the eavesdropping link is given by
\begin{align}\label{sinr for ES}
\text{SINR}_{kj} & = \frac{\gamma P_{\rm rec}\left(d_{kj,0}\right)}{(1-\gamma) P_{\rm rec}\left(d_{kj,0}\right)+\mathcal{I}_{kj}+\sigma_{\rm S}^2},
\end{align}
where \(P_{\rm rec}(d_{kj, i})\) indicates the received power at the eavesdropper from both typical (\(i=0\)) and interfering (\(i\neq 0\)) \ac{IoT} devices. 
\begin{equation}
\mathcal{I}_{kj}=\sum\limits_{i\in \Phi_u\setminus \{i=0\}}  P_{\rm rec}(d_{kj, i})
\end{equation}
represents the total interference from other \ac{IoT} devices (\(i\neq 0\)) at the \ac{ES}.

\par
In conclusion, the addition of \ac{AN} considerably impairs the eavesdropper's ability to intercept the communication while leaving the reception for the intended receiver unaffected. By carefully managing the power of the \ac{AN}, it is feasible to markedly improve the confidentiality of uplink transmissions from \ac{IoT} devices to \ac{LEO} satellites, thereby enhancing security at the physical layer.

\section{Performance Analysis}
In secure satellite communications, evaluating the effectiveness of \ac{PLS} techniques requires a comprehensive understanding of how well the system can maintain connectivity, protect against eavesdropping, and ensure reliable and secure data transmission. The performance metrics, availability probability, coverage probability, and secure connection probability are integral to assessing these aspects. The multi-tier architecture is central to our analysis, enabling us to model various orbital configurations and assess their impacts on key performance metrics. This approach allows evaluation under different conditions, including potential eavesdropping scenarios, highlighting its significance in optimizing secure communication in \ac{LEO} satellite-enabled \ac{IoT} networks.  In this section, we evaluate the security of the considered system network using the above performance metrics or their combination and provide analytical expressions for these performance metrics.

\par To facilitate analysis, we begin by characterizing the distribution of the central angle from the typical \ac{IoT} device to the serving satellite, which is known as the contact angle \cite{9861378}. Given that $N_m$ satellites are modeled uniformly as \ac{BPP}, it follows that any randomly chosen points share an identical distribution of the contact angle. Therefore, the \ac{CDF} and \ac{PDF} of the contact distance can be articulated in the following lemma.

\begin{lemma}[Contact Angle Distribution]\label{lemma:pdf contact angle}
Given the typical \ac{IoT} device and the closest satellite within the $m$-th tier, the \ac{CDF} of the contact angle $\theta_{m,0}$ is given by
\begin{equation}\label{eq:cdf_contact_angle}
     F_{\theta_{m,0}}(\theta) = 1- \left(\frac{1 + \cos{\theta}}{2}\right)^{N_m}, \quad 0 < \theta < \theta_{m, \mathrm{max}},
\end{equation}
and the corresponding \ac{PDF} is 
\begin{equation}\label{eq:pdf_contact_angle}
     f_{\theta_{m,0}}(\theta) = \frac{N_m \sin{\theta}}{2} \left(\frac{1 + \cos{\theta}}{2}\right)^{N_m-1}, \quad 0 < \theta < \theta_{m, \mathrm{max}},
\end{equation}
where $\theta_{m, \mathrm{max}}$ is defined in (\ref{eq: theta_max}). 
\begin{IEEEproof}
See appendix~\ref{Proof: pdf contact angle}.
\end{IEEEproof}
\end{lemma}
Next, we derive the availability probability, which is directly related to the contact angle distribution. 
The availability probability stands as a fundamental metric, establishing the foundational viability of any communication attempt.  Without satellite availability, neither data transmission nor secure communication can occur, as defined and mathematically presented as follows: 

\begin{definition}\label{Def Avai Prob}
Availability probability is the probability that, within the visible area of a typical \ac{IoT} device, at least one satellite from the $m$-th tier is available for communication. It is mathematically presented as follows:
\begin{equation}
     \mathcal{P}_{m, \rm av} \triangleq \mathbb{P} \left[\mathcal{N}\left(\mathcal{A}_{m, \rm vis}\right) > 0\right],
\end{equation}
where \(\mathcal{A}_{m, \rm vis}\) is  the surface area from which satellites at tier  $m$  are visible to the typical \ac{IoT} device, defined by the set $\mathcal{A}_{m, \rm vis}= \{ R_m, 0\leq \theta\leq  \theta_{m,\rm max}, 0\leq \varphi \leq 2\pi\}$. $\mathcal{N}(\mathcal{A}_{m, \rm vis})$ counts the number of satellites in the surface area $\mathcal{A}_{m, \rm vis}$.
\end{definition}
Next, we derive the availability probability in the following lemma as defined in Definition \ref{Def Avai Prob}, which is directly related to the contact angle distribution.

\begin{lemma}[Availability Probability]\label{lemma:avai prob eq}
The probability of availability $\mathcal{P}_{m, \mathrm{av}}$ of a satellite in the $m$-th tier is given as 
\begin{align}\label{eq of avai prob}
    \mathcal{P}_{m, \rm av}=1-\left(\frac{1+\cos{\theta_{m,\rm max}}}{2}\right)^{N_m}, 
\end{align} 
where  $\theta_{m,\rm max}$ is defined in (\ref{eq: theta_max}) and $N_m$ is the total number of satellites within tier $m$. 
\begin{IEEEproof}
Essentially, the \ac{CDF}  of contact angle distribution, which is defined as $\mathbb{P}\left[\theta_{m,0} < \theta\right]$ is equivalent to the availability probability
$\mathbb{P} \left[\mathcal{N}\left(\mathcal{A}_{m, \rm vis}\right)=0\right]$, when $\theta_{m,0}=\theta_{m, \rm max}$, where $\mathcal{A}_{m, \rm vis}$ describes the maximum visibility area. Following the void probability, the lemma is proved.
\end{IEEEproof}
\end{lemma}
Once the availability of satellites is determined,  the analysis progresses to the coverage probability. Coverage probability assesses the likelihood that a communication link, specifically between a typical \ac{IoT} device and its serving satellite, can be established with sufficient signal quality. This quality is necessary to decode transmitted signals with minimal errors accurately. It is conditional upon the \ac{SINR} exceeding a predetermined threshold, $\beta_{\rm LS}$ and is defined as follows:  
\begin{definition}\label{Def Cov Prob}
The coverage probability is the probability that the \ac{SINR} for the legitimate link is greater than $\beta_{\rm LS}$, thereby ensuring reliable signal decoding. Formally, it is expressed as:
\begin{align}
     \mathcal{P}_{m, \rm cov}&\triangleq   \mathbb{P}\left[{\rm SINR}_{\rm LS}>\beta_{\rm LS}\right],
\end{align}
where the ${\rm SINR}_{\rm LS}$ is defined in (\ref{sinr for LS}).
\end{definition} 
Prior to the analysis of coverage probability, we need to understand the impact of interference on the communication link.  Given the spatial distribution of \ac{IoT} devices follows \ac{PPP}, we employ the \ac{LT} to obtain more manageable and closed-form expressions for interference analysis. 

\begin{lemma}[Laplace Transform of Interference]\label{lemma:LT}
For a given satellite located in tier $m$, the \ac{LT} of the interference \(\mathcal{L}_{\mathcal{I}_m}(s)\) experienced by this satellite is:
\begin{align}
\mathcal{L}_{\mathcal{I}_m}\left(s\right) = \exp &\left(-  \lambda_u \int\limits_0^{2\pi}\int\limits_0^{ \theta_{m, \rm max}}\left[1-\left(1+m_2 s P_t G \left(\frac{c}{4 \pi f d}\right)^2 \right)^{-m_1}\right] \right. \nonumber \left. \vphantom{\int\limits_0^{2\pi}} R_\oplus^2 \sin{\theta}  d\theta d\varphi\right),
\end{align}
where $\theta_{m, \rm max}$ is defined in (\ref{eq: theta_max}).
\begin{IEEEproof}
See appendix \ref{Proof: lemma of LT}.
\end{IEEEproof}
\end{lemma}
Building upon the interference analysis, the derivation of the coverage probability proceeds as follows.

\begin{lemma}[Coverage Probability]\label{lemma cov prob} 
The coverage probability for the uplink communication between a typical \ac{IoT} device and its nearest \ac{LS} in the $m$-th tier with SINR threshold $\beta_{\rm LS}$ is derived as follows: 
    \begin{align}
    \mathcal{P}_{m, \rm cov}= \sum\limits_{q=1}^{m_1}\binom{m_1}{q} (-1)^{q+1}\int\limits_0^{\theta_{m,\rm max}} \exp{\left( -q s_{\rm LS}\sigma_S^2\right)}&\mathcal{L}_{\mathcal{I}_{m}}\left(q s_{\rm LS}\right)  \nonumber f_{\theta_{m,0}}(\theta)d\theta,
\end{align}
where $s_{\rm LS}=\frac{(m_1!)^{-\frac{1}{m_1}}}{m_2}\frac{\beta_{\rm LS}}{ \gamma P_t G}\left(\frac{4 \pi f d_{m,0}}{c}\right)^2$,  and  $\theta_{m, \rm max}$ is as defined in (\ref{eq: theta_max}). The functions \( f_{\theta_{m,0}}(\cdot)\), $\mathcal{P}_{m, \rm av}$,  and $\mathcal{L}_{\mathcal{I}_{m}}(\cdot)$ are provided in Lemma \ref{lemma:pdf contact angle}, Lemma \ref{lemma:avai prob eq} and Lemma \ref{lemma:LT}, respectively.
\begin{IEEEproof}
See appendix \ref{Proof: lemma cov prob}.
\end{IEEEproof}
\end{lemma}
We are ready to introduce the concept of successful communication probability, one of the main metrics, which involves the insights gained from the availability and coverage probabilities. 
The successful communication probability denoted as $\mathcal{P}_{m, \rm suc}$,  is defined as the probability that an uplink connection made by the typical \ac{IoT} device to its nearest satellite in the  $m$-th tier successfully meets both availability and coverage criteria. Mathematically, this probability is expressed as the product of the availability probability and the coverage probability:
\begin{align}\label{thm suc prob}
    \mathcal{P}_{m, \rm suc}= \mathcal{P}_{m, \rm av} \times \mathcal{P}_{m, \rm cov},
\end{align}
where $\mathcal{P}_{m, \rm av}$  and $\mathcal{P}_{m, \rm cov}$ are derived in Lemma \ref{lemma:avai prob eq}  and Lemma \ref{lemma cov prob}, respectively.
\par Moving beyond the feasibility of communication, the definition of secrecy outage probability addresses the security aspect, which investigates the scenario where the \ac{SINR} experienced by any potential  \ac{ES} falls below a certain predefined threshold, $\beta_{\rm ES}$. Hence, within a multi-tiered satellite constellation framework, we define and derive the secrecy outage probability as follows.
\begin{definition}\label{Def Sec Prob}
Secrecy outage probability is the probability that the highest \ac{SINR} experienced by any \ac{ES} across a multi-tiered satellite constellation falls below a predefined threshold \(\beta_{\rm ES}\). Mathematically, $\mathcal{P}_{\rm out}$ is defined as:
\begin{align}
     \mathcal{P}_{\rm out }&\triangleq  \mathbb{P}\left[\max_{k\neq m} \max_{j\in \Phi_{\rm S_k}}\left\{ \text{\rm SINR}_{kj}\right\}<\beta_{\rm ES} \right], 
\end{align} 
where ${\rm SINR}_{kj}$ represents the \ac{SINR} experienced by $j$-th  \ac{ES} within the $k$-th tier, excluding the legitimate $m$-th tier , where the expression is given in  (\ref{sinr for ES}). 
\end{definition}
Upon the definition of the secrecy outage probability, we can derive the expression of the $ \mathcal{P}_{\rm out }$ in the following lemma.

\begin{theorem}[Secrecy Outage Probability]\label{lemma out prob}
For a given value of $\beta_{\rm ES}$, the probability of secrecy outage, as defined in Definition \ref{Def Sec Prob}, is derived as:
\begin{align}\label{eq of out prob}
\mathcal{P}_{\rm out} = \prod_{\substack{k=1 \\ k \neq m}}^{K}   & \left[\sum\limits_{q=0}^{m_1}\binom{m_1}{q} (-1)^q  \int\limits_0^{\theta_{k, \rm max}}\exp{\left(-q s_{\rm ES}\sigma_S^2\right)}\mathcal{L}_{\mathcal{I}_{kj}}\left(q s_{\rm ES}\right) \right. \nonumber \left. \vphantom{\int\limits_0^{2\pi}} \times f_{\theta_{k,0}}(\theta)d\theta +\frac{1+\cos{\theta_{k, \rm max}}}{2}\right]^{N_k},
\end{align}
with  \(s_{\rm ES}=\frac{(m_1!)^{-\frac{1}{m_1}}}{m_2}\frac{\beta_{\rm ES}}{\left[\gamma -\beta_{\rm ES} (1-\gamma) \right]P_t G}\left(\frac{4 \pi f d_{kj,0}}{c}\right)^2\) and $f_{\theta_{k,0}}(\theta)=\frac{\sin{\theta}}{2}$
The \ac{LT} of interference, \(\mathcal{L}_{\mathcal{I}_{kj}}(\cdot)\), is derived in Lemma \ref{lemma:LT}. 
\begin{IEEEproof}
See appendix \ref{Proof: Lemma Out Prob}.
\end{IEEEproof}
\end{theorem}
Finally, our contribution goes further by not only considering the successful connection but also ensuring the data transmitted is secure from eavesdropping. The secure communication probability, denoted as $\mathcal{P}_{\rm sec}$, is calculated as the product of two key factors: the probability of the typical \ac{IoT} device achieving successful communication within the legitimate $m$-th tier and the secrecy outage probability considering the influence of all  \ac{ES}. Mathematically, the secure communication probability is formulated as:
\begin{align}
\mathcal{P}_{\rm sec}=   \mathcal{P}_{m, \rm suc} \times \mathcal{P}_{\rm out},
\end{align}
where $\mathcal{P}_{\rm out}$, the secrecy outage probability is elaborated on in Theorem~\ref{lemma out prob}, and $ \mathcal{P}_{m, \rm suc}$, the probability of achieving successful communication, is described in (\ref{thm suc prob}).

\begin{figure*}[!ht]
\centering
\subfigure[$\mathcal{P}_{m, \rm av}$ against beamwidth angle and  the altitude of satellites]{\includegraphics[width=0.49\textwidth]{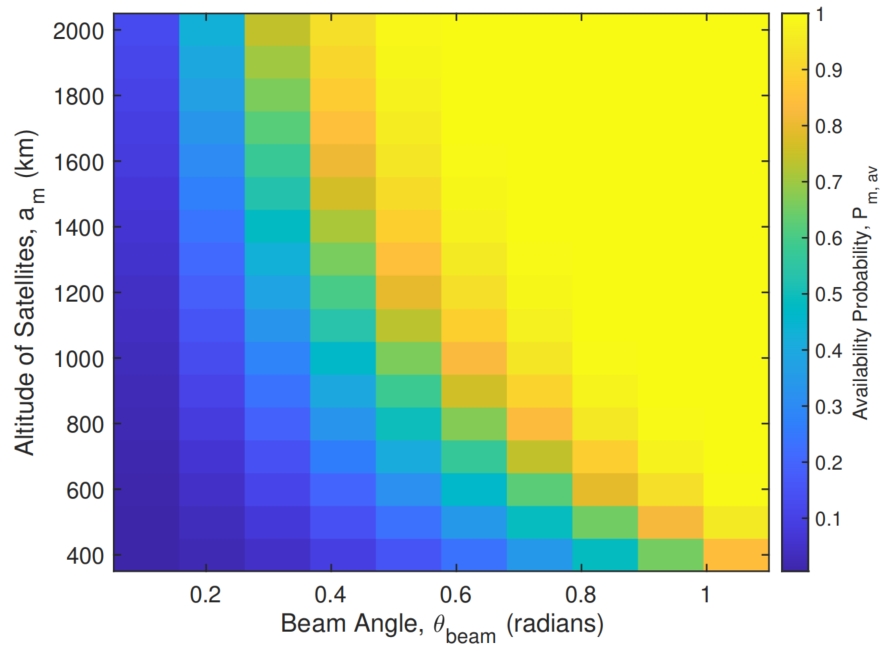}\label{Feb_13_avai_alt_beam_1b}}
\subfigure[$\mathcal{P}_{m, \rm av}$ against number of satellites and the altitude of satellites]{\includegraphics[width=0.49\textwidth]{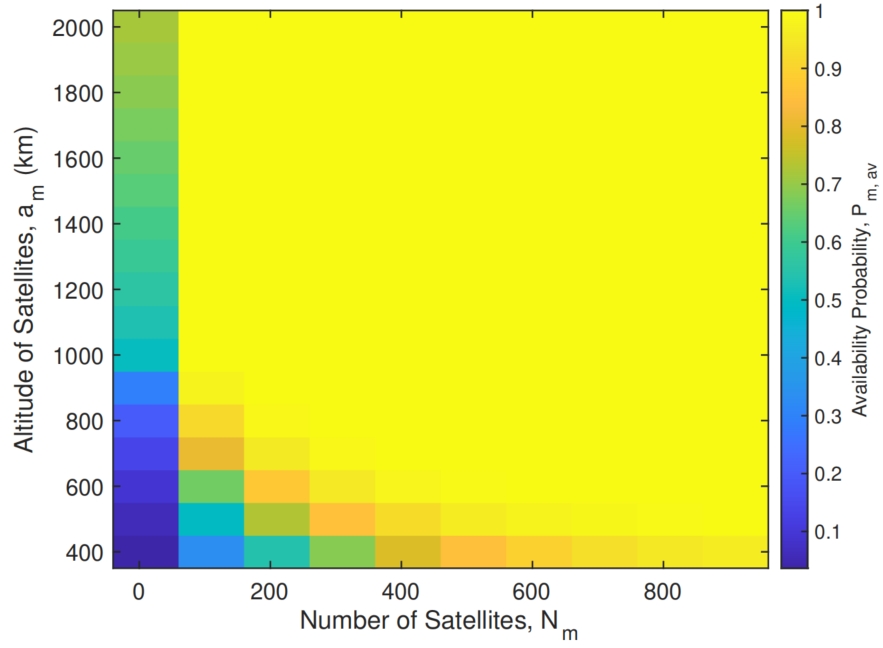}\label{Feb_13_avai_sat_alt_1a}}
\caption{Heat map representing the $ \mathcal{P}_{m, \rm av}$ for different parameters of satellite network.}
\label{Avai_Prob}
\end{figure*}

\begin{table}[!t] 
\centering
\caption{Simulation Parameters} 
\label{Table of simulation}
\resizebox{0.75\linewidth}{!}{\begin{tabular}{c|c|c}
\hline \textbf{Notation} & \textbf{Parameter} &  \textbf{Default Value} \\ \hline 
\hline$f_c$ &   Center frequency & 2 $\mathrm{GHz}$ \\
\hline $R_{\oplus} $ & Earth radius & 6371 km \\
\hline $N_k$ & Number of satellites in tier $k$, $\forall k$ & 500\\
\hline $a_k$ &  Altitude of satellites for all tier, $k=1,2,3$  &  [500, 1000, 1500] km \\
\hline $\theta_{\rm beam}$ & Half beamwidth angle of satellite & ${\pi}/{3}$\\
\hline $P_t$  &  Uplink transmit power of \ac{IoT} devices & 23 $\mathrm{dBm} $ \\ 
\hline $\lambda_u$ &  Density of users (IoT devices) &  $10^{-6}$ devices$/$km$^{-2}$\\
\hline $N_0$ & Noise spectral density  & -174 $\mathrm{dBm/Hz}$ \\ 
\hline $B$ & Bandwidth & 180 $\mathrm{KHz}$ \\ 
\hline $\Gamma(m_1,m_2)$ & Gamma shape and scale parameters  & $[m_1,m_2]=[1,0.1269]$ \\
\hline $\gamma$ &  AN power allocation coefficient &  0.1 \\
\hline $G$ & Antenna  bea, gain  &  41.9  $\mathrm{dBi}$\\
\hline $\beta_{\rm ES}$ & \ac{SINR}  threshold at \ac{ES}  &  -10 $\mathrm{dB}$\\
\hline $\beta_{\rm LS}$ & \ac{SINR} threshold at \ac{LS}  & -30 $\mathrm{dB}$ \\
\hline
\hline
\end{tabular}}
\end{table}

\section{Numerical Results}
In this section, we present numerical results for the derived performance metrics. We employed Monte Carlo simulations with 10,000 iterations. In each iteration, the positions of the satellites and \ac{IoT} devices were varied to simulate the dynamic changes in the topology. This approach ensures that the performance metrics derived from our analytical expressions correspond to the average performance of the network under dynamic conditions rather than a static snapshot. We use lines to present analytical results and markers to present Monte Carlo simulation results. The matching between them in the figures provided in this section aims to affirm the accuracy of our analysis further. The parameters employed in these simulations are detailed in Table~\ref{Table of simulation}. 
\subsection{System Design on Network Availability}
This subsection provides how the design of satellite networks influences a network's ability to establish communication links between \ac{IoT} devices and satellites. As shown in (\ref{eq of avai prob}), the availability probability $\mathcal{P}_{m, \rm av}$ is influenced by the beamwidth, number, and altitude of satellites. In Fig.~\ref{Avai_Prob}, we highlight the optimal parameters design with regard to $\mathcal{P}_{m, \rm av}$.

\par
In Fig.~\ref{Feb_13_avai_alt_beam_1b}, as the beamwidth angle increases, the area covered by each satellite's communication beam expands, increasing the availability probability $\mathcal{P}_{m, \rm av}$. Furthermore, with the same beam angle, satellites at higher altitudes cover a larger area of the Earth's surface, thus increasing $\mathcal{P}_{m, \rm av}$. With a fixed constellation size (e.g., $N_m=500$), we observe that we can reach $\mathcal{P}_{m, \rm av}\approx 1$  for satellite altitudes higher than 500 km when the beamwidth angle is set to $\theta_{\rm beam}=\frac{\pi}{3}$. 

\par
Fig.~\ref{Feb_13_avai_sat_alt_1a} demonstrates how $\mathcal{P}_{m, \rm av}$ is influenced by constellation configurations. Increasing the number and altitude of satellites can enhance $\mathcal{P}_{m, \rm av}$. The analysis identifies a constellation size of $N_m=500$ satellites as particularly effective in securing near-universal availability across all network tiers with $\mathcal{P}_{m, \rm av} \approx 1$. Further increasing the number of satellites offers limited benefits to availability.

\begin{figure}[!ht]
\centering
\subfigure[$\theta_{\rm beam}= \pi / 3$]{\includegraphics[width=0.49\textwidth]{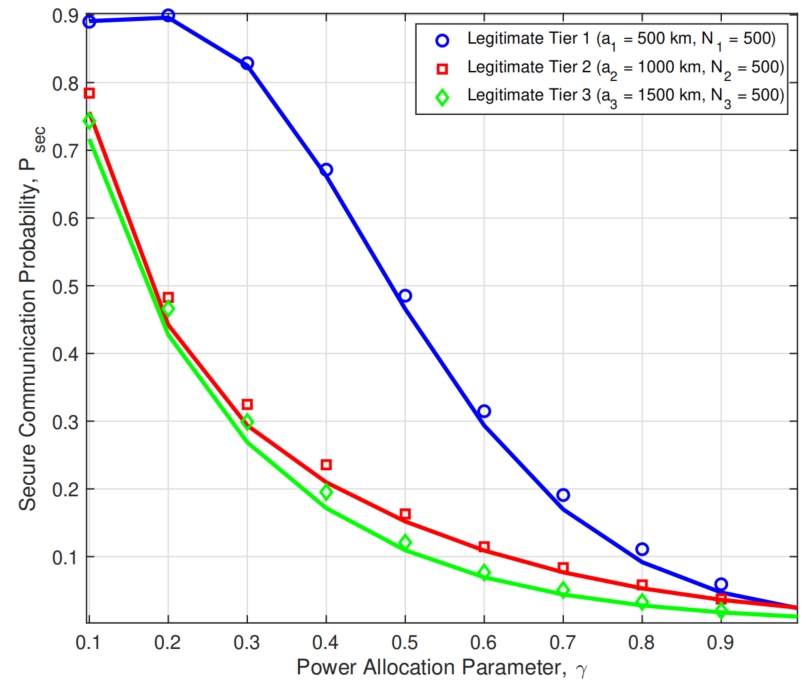}\label{Feb_20_sec_an_1_EXT}}
\subfigure[$\theta_{\rm beam}= \pi / 3$]{\includegraphics[width=0.49\textwidth]{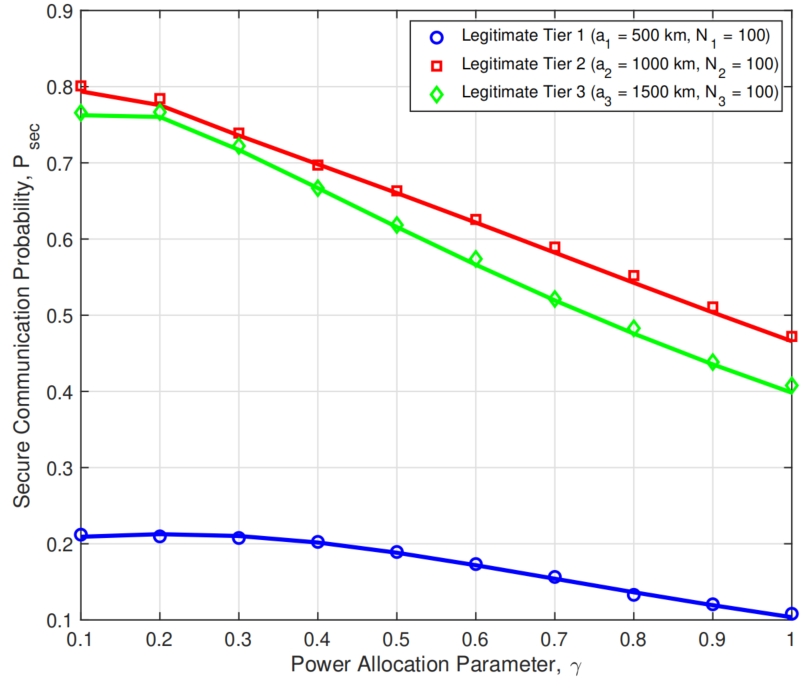}\label{Feb_20_sec_an_2_EXT}}
\subfigure[$\theta_{\rm beam}= \pi / 4$]{\includegraphics[width=0.49\textwidth]{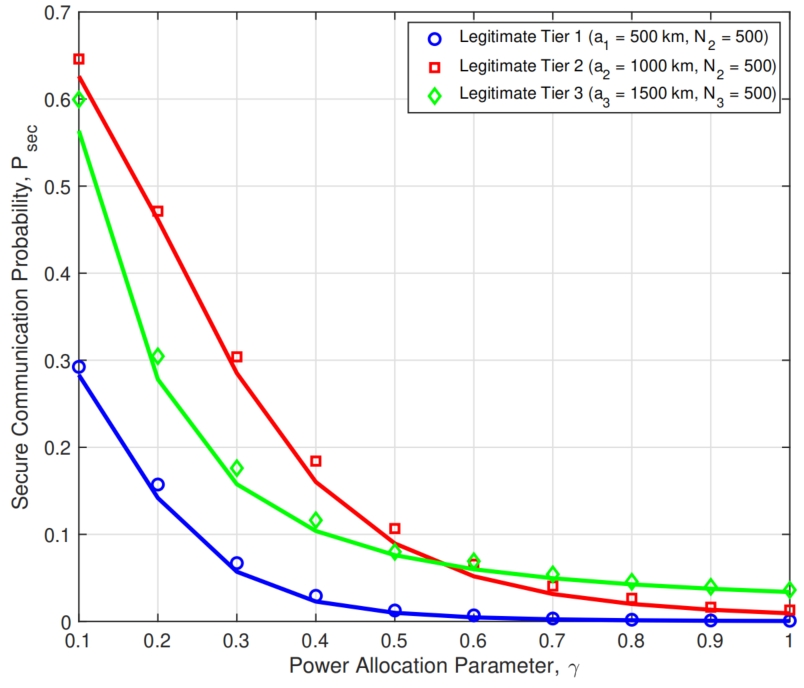}\label{Feb_20_sec_an_3_EXT}}
\caption{The effects of power allocation across three configurations on $\mathcal{P}_{\rm sec}$.}
\label{Sec_Prob_4}
\end{figure}

\begin{figure}[!ht]
\centering
\includegraphics[width=0.6\textwidth]{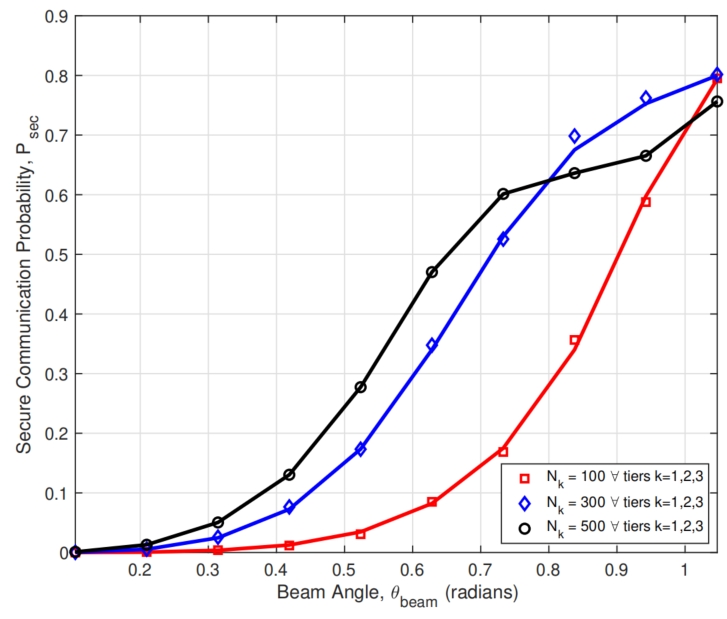}
\caption{Impact of satellite numbers on $\mathcal{P}_{\rm sec}$ against beam angle.}
\label{Mar_5_sec_beam_1a_EXT}
\end{figure}

\subsection{Enhancing PLS: Techniques and Trade-offs}
This subsection discusses how different strategies can influence the secure communication probability $\mathcal{P}_{\rm sec}$.
Unless otherwise stated, we adopt a 3-tier constellation, with altitudes set at $a_{\rm S}=[500, 1000, 1500]$ km. The number of satellites for all tiers is fixed as $N_k=500$, $\forall k=1,2,3$. 
\par Fig.~\ref{Sec_Prob_4} shows how different choices of the legitimate tier and the power allocation parameter $\gamma$ influence $\mathcal{P}_{\rm sec}$.  We observe a consistent trend across all subfigures: $\mathcal{P}_{\rm sec}$ decreases as $\gamma$ increases. This is because less power is allocated to \ac{AN} for a higher $\gamma$, making it easier for eavesdroppers to decode the signal. More precisely, the transition from Fig.~\ref{Feb_20_sec_an_1_EXT} to Fig.~\ref{Feb_20_sec_an_2_EXT} and Fig.~\ref{Feb_20_sec_an_3_EXT} illustrates changes in satellite number and beam angle. When the satellite count is increased, the chance of having a satellite available for communication  ($\mathcal{P}_{m, \rm av}$) is high, leading to Tier-1 having the highest $\mathcal{P}_{\rm sec}$ as depicted in Fig.~\ref{Feb_20_sec_an_1_EXT}. However, more satellites can also create more interference, particularly at higher altitudes, thus lowering  $\mathcal{P}_{\rm sec}$ in Tiers 2 and 3. Conversely, in Fig.~\ref{Feb_20_sec_an_2_EXT}, reducing the number of satellites significantly impacts Tier-1, making it less secure due to a lower $\mathcal{P}_{m, \rm av}$; thus, Tier-2 becomes the preferred choice for the legitimate tier due to its balanced performance. 
Fig.~\ref{Feb_20_sec_an_3_EXT} shows the effects of reducing beam angle from $\theta_{\rm beam}=\frac{\pi}{3}$ to $\theta_{\rm beam}=\frac{\pi}{4}$ while maintaining $N_k=500$, $\forall k$. In this configuration, Tiers 2 and 3 outperform Tier 1 across the entire range of the $\gamma$. The narrower beam angle implies a need for more satellites to maintain high $\mathcal{P}_{m, \rm av}$. Overall, we can draw the same conclusion because whether reducing $N_k$ $\forall k$, or $\theta_{\rm beam}$ decreases the $\mathcal{P}_{m, \rm av}$.
\par 

\par
\begin{figure}[!ht]
\centering
\includegraphics[width=0.6\textwidth]{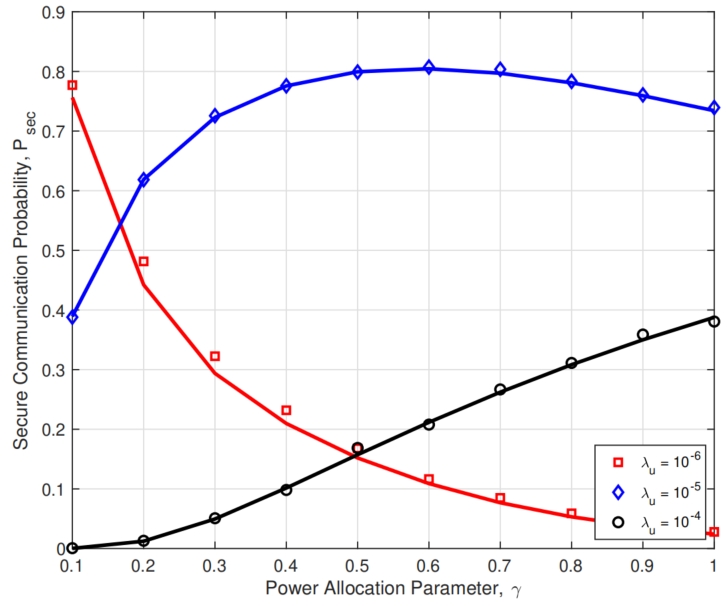}
\caption{$\mathcal{P}_{\rm sec}$ as a function of power allocation parameter across different densities of IoT devices.}
\label{Feb_23_sec_an_1_EXT}
\end{figure}
\begin{figure*}[!ht]
\centering
\subfigure[$\lambda_u=10^{-6}$]{\includegraphics[width=0.49\textwidth]{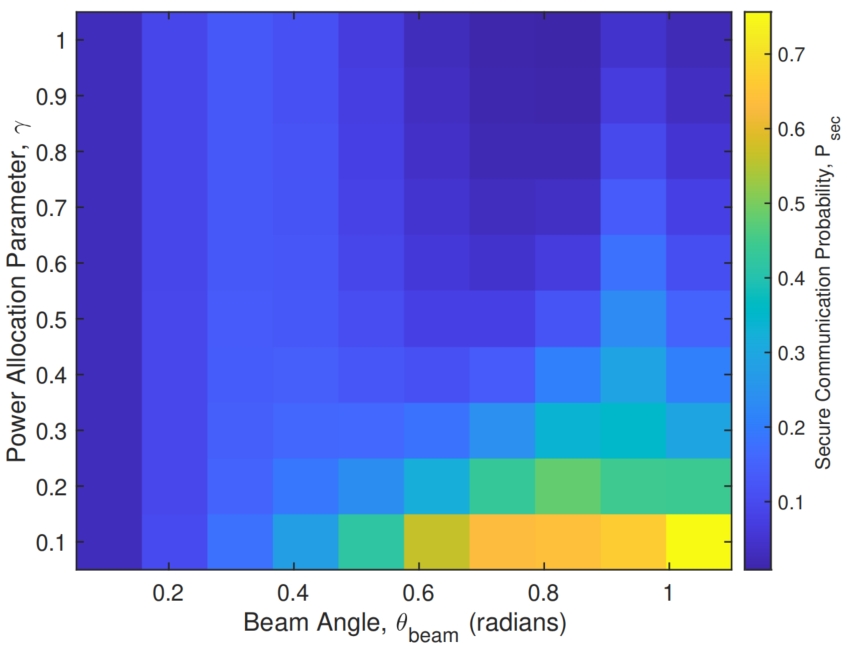}\label{Mar_5_hm_beam_an_1b}}
\subfigure[$\lambda_u=10^{-5}$]{\includegraphics[width=0.49\textwidth]{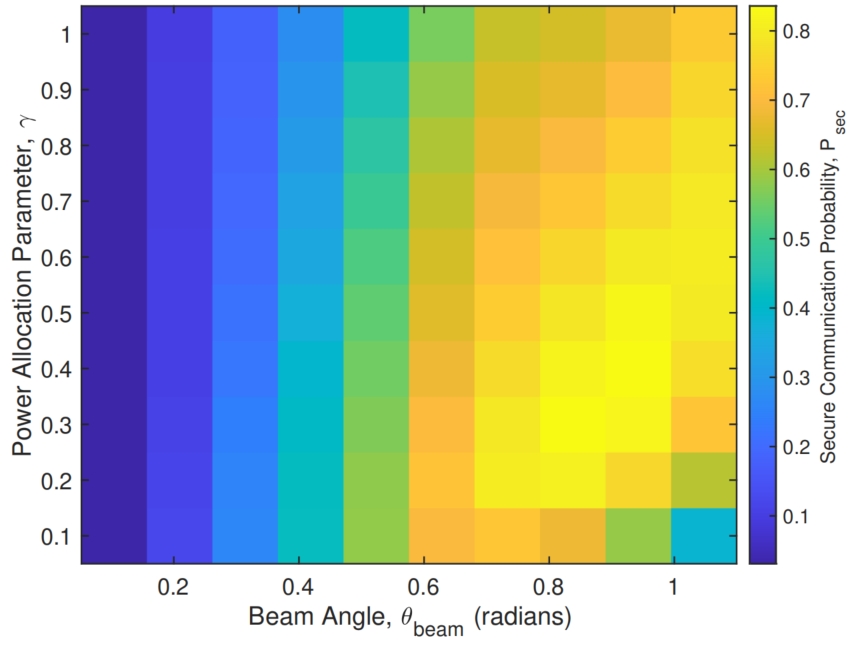}\label{Mar_5_hm_beam_an_1c}}
\caption{Optimal beam angle and AN  power allocation for maximizing $\mathcal{P}_{\rm sec}$ in different IoT devices.}
\label{Sec_Prob_5}
\end{figure*}

The adjustments illustrate a trade-off between availability and signal quality, indicating that Tier-2, with an altitude of $a_m=1000$ km, is the optimal choice for the legitimate tier due to its consistently stable maximum $\mathcal{P}_{\rm sec}$ as depicted in Fig.~\ref{Sec_Prob_4}. Therefore, unless otherwise stated, subsequent analyses of system performance in Figs.~\ref{Mar_5_sec_beam_1a_EXT}, \ref{Feb_23_sec_an_1_EXT}, and \ref{Sec_Prob_5} focus on Tier-2 as the legitimate tier.
\par 
Fig.~\ref{Mar_5_sec_beam_1a_EXT} highlights the nuanced impact of varying satellite numbers and beam angles on $\mathcal{P}_{\rm sec}$.  A key trade-off emerges in the number of satellites: more satellites can improve coverage and security up to a point, but beyond that, they might introduce excessive interference or potential eavesdroppers. In this case, we should emphasize the need for an optimized satellite constellation design.

\par Moving to Fig.~\ref{Feb_23_sec_an_1_EXT}, as  $\gamma$ increases, the behaviour of $\mathcal{P}_{\rm sec}$ is totally differently for the three densities  of \ac{IoT} devices. For the sparsest \ac{IoT} network ($\lambda_u=10^{-6}$), $\mathcal{P}_{\rm sec}$ exhibits a peak at lower $\gamma$ values. When interference is negligible, allocating more power to \ac{AN}  can effectively protect the \ac{PLS}. In contrast, in the high-density scenario ($\lambda_u=10^{-4}$), an inverse trend is observed where $\mathcal{P}_{\rm sec}$ is the lowest at $\gamma=0.1$ but increases steadily. This trend reflects that ensuring the information can effectively reach \ac{LS}s in an interference-limited scenario is more important than preventing eavesdropping. When $\lambda_u=10^{-5}$, the trend of $\mathcal{P}_{\rm sec}$ initially rising and then declining suggests that balancing between ensuring successful, legitimate communication and preventing eavesdropping is crucial. In summary, Fig.~\ref{Feb_23_sec_an_1_EXT} highlights a critical density-dependent optimization strategy for allocating power between messaging and \ac{AN}  to maximize the $\mathcal{P}_{\rm sec}$.

\begin{figure*}[!ht]
\centering
\subfigure{\includegraphics[width=0.49\textwidth]{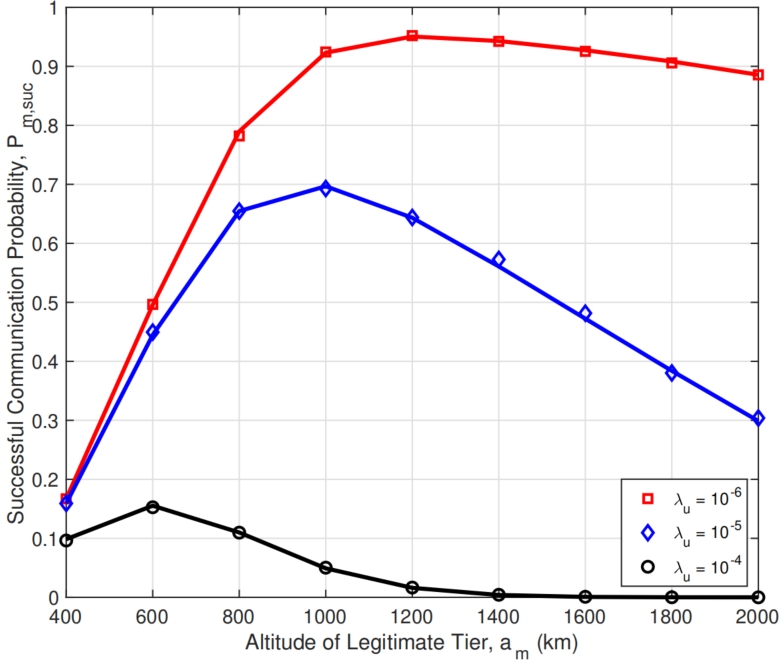}\label{Feb_23_sec_alt_4_cov}}
\subfigure{\includegraphics[width=0.49\textwidth]{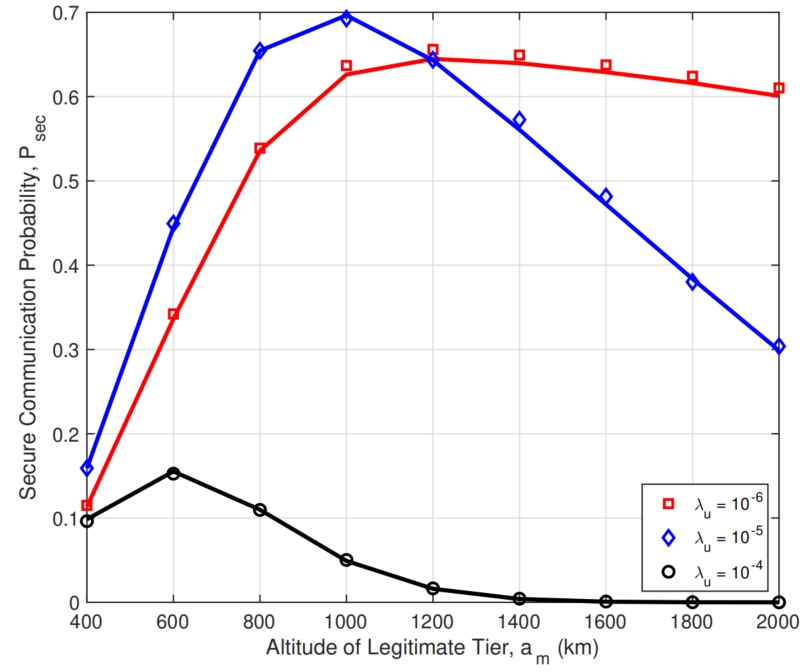}\label{Feb_23_sec_alt_4_sec}}
\caption{Impact of the LS altitude on $\mathcal{P}_{m, \rm suc}$ and $\mathcal{P}_{\rm sec}$ across different IoT device densities.}
\label{Feb_23_sec}
\end{figure*}


\subsection{Security Performance Optimization}
In this subsection, we explore adjustments to the network configuration and security outcomes to maximize or achieve optimal security.
\par Fig.~\ref{Sec_Prob_5}  visually represents the optimization of $\mathcal{P}_{\rm sec}$ in sparse ($\lambda_u=10^{-6}$) versus denser ($\lambda_u=10^{-5}$) \ac{IoT} devices deployment. There is a clear distinction in how $\theta_{\rm beam}$ and $\gamma$ influence $\mathcal{P}_{\rm sec}$ at different device densities.
As shown in Fig.~\ref{Mar_5_hm_beam_an_1b}, for lower density ($\lambda_u=10^{-6}$), $\mathcal{P}_{\rm sec}$ is more sensitive to the configuration of $\theta_{\rm beam}$ and $\gamma$, favoring wider beam angles and a large share of power dedicated to AN. This indicates that securing communications relies more on \ac{AN}  in environments with fewer devices, with a wider signal spread proving beneficial due to the lower interference risk.
Conversely, Fig.~\ref{Mar_5_hm_beam_an_1c} corresponds to a denser network ($\lambda_u=10^{-5}$), shows a different trend, where $\mathcal{P}_{\rm sec}$   demonstrates reduced sensitivity to variations in $\gamma$, especially at lower beam angles. The optimal $\mathcal{P}_{\rm sec}$ levels, as indicated by a yellow gradient, are observed across a broader range of beam angles compared to Fig.~\ref{Mar_5_hm_beam_an_1b}. This pattern suggests that high-security levels are attainable with a variety of $\gamma$ settings, with the yellow regions extending throughout the full range of $\gamma$ values from low to high.


Overall, lower-density environments show the highest secrecy with significant AN, while higher-density environments maintain high secrecy across varying levels of \ac{AN} power.

\par In Fig.~\ref{Feb_23_sec}, we explore the effects of varying altitudes in the legitimate tier while keeping the altitudes of the potential eavesdropping tiers fixed at $500$ and $1500$ km. The optimal altitude exists for both $\mathcal{P}_{m,\rm suc}$ and $\mathcal{P}_{\rm sec}$  across varying densities of \ac{IoT} devices. Specifically, the optimal legitimate altitude decreases as the density increases. At very high densities, the optimal altitude is found to be 600 km. As the density decreases slightly, this optimal altitude increases to 1000 km, and with a further reduction in density, it reaches 1200 km. This is the result of a trade-off between availability and received signal strength. In most cases, as the density of \ac{IoT} devices increases, $\mathcal{P}_{m,\rm suc}$ and $\mathcal{P}_{\rm sec}$ tends to decrease. When $\lambda_u=10^{-5}$, $\mathcal{P}_{\rm sec}$ presents  more nuanced results.  We interpret this interesting phenomenon as interference protecting signal security. This protective mechanism is similar to \ac{AN}  technology, where the received \ac{SINR} of \ac{LS}s is sacrificed to introduce more interference to eavesdroppers. However, this trade-off has its limits, as demonstrated by the highest density ($\lambda_u=10^{-4}$), which
does not yield any beneficial impact on $\mathcal{P}_{\rm sec}$. This suggests a threshold beyond which additional interference no longer
benefits securing communications.

\section{Conclusion}
In conclusion, this paper represents a significant advancement in securing \ac{IoT} communications via \ac{LEO} satellites by implementing \ac{PLS} strategies enhanced by \ac{SG} tools and \ac{AN} techniques. The primary contribution of this study lies in its innovative approach to modeling the spatial distribution of legitimate and  \ac{ES}s as well as \ac{IoT} devices. This enables a more accurate analysis of security threats and the effectiveness of \ac{PLS} measures within a \ac{LEO} satellite-enabled \ac{IoT} network. By integrating \ac{AN} into the \ac{PLS} framework and utilizing \ac{SG} for network analysis, this research not only addresses the substantial security challenges presented by the extensive coverage and accessibility of \ac{LEO} satellites but also provides a scalable solution capable of adapting to the dynamic nature of satellite positions and fluctuating channel conditions. However, the trade-off analysis suggests that while the proposed model significantly enhances security, careful management of \ac{AN} is required to prevent compromising communication integrity. Furthermore, the model operates under the assumption of known eavesdropper altitudes, which may not always be applicable in practical scenarios. Nonetheless, these findings offer valuable insights into ongoing efforts to secure satellite-based \ac{IoT} networks, proposing avenues for future research to enhance the robustness and applicability of \ac{PLS} solutions in such complex and evolving communication environments.

\appendices
\section{Proof of Lemma \ref{lemma:pdf contact angle}}\label{Proof: pdf contact angle}
We consider the spatial distribution of satellites within a homogeneous \ac{BPP} across the surface of a sphere; we focus on the $m$-th tier (i.e., the sphere $S_m$) to ascertain the distribution characteristics of the nearest satellite. Within this framework, the probability $\mathbb{P}[\theta_{m,0} \leq \theta]$ that a randomly chosen satellite falls within a spherical cap defined by a central angle $\theta$ is proportional to the area of this cap relative to the sphere's total surface area. Mathematically, this probability is given by:
\begin{align*}
    \mathbb{P}[\theta_{m,0} \leq \theta]
    & =\frac{\mathcal{S} \left(\mathcal{A}_{m, \rm vis}\right)}{\mathcal{S} \left(\mathcal{A}_{S_m}\right)}= \frac{2\pi R_m^2(1 - \cos \theta)}{4\pi R_m^2}.
\end{align*}
Simplifying, we find:
\[
\mathbb{P}[\theta_{m,0} \leq \theta] = \frac{1 - \cos \theta}{2}.
\]

The \ac{CDF}  for the contact angle $F_{\theta_{m,0}}(\theta)$, considers the case where the serving satellite is the nearest among all $N_m$ \ac{i.i.d.} satellites and derived easily as follows
\begin{align*}
    F_{\theta_{m,0}}(\theta)&=1-\prod\limits_1^{N_m}  \mathbb{P}[\theta_{m,0} > \theta]=1 - \left[1 - \frac{1 - \cos \theta}{2}\right]^{N_m}.
\end{align*}
This equation captures the probability that at least one satellite is within the spherical cap, effectively accounting for the presence of $N_m$ satellites in the calculation. The \ac{PDF} can be obtained by differentiating the \ac{CDF}  with respect to $\theta$. 

\section{Proof of Lemma \ref{lemma:LT}}\label{Proof: lemma of LT}
Recall that the interference is expressed as
\begin{equation}
\mathcal{I}_m=\sum\limits_{i\in \Phi_u\setminus \{i=0\}}  P_t \left(\frac{c}{4 \pi f d_{m,i}}\right)^2 G |h_{m,i}|^2,
\end{equation}
where the $d_{m,i}$ is the distance between the serving LS with the interfering $i$-th \ac{IoT} device.
We proceed with the derivation of the \ac{LT} of interference by definition:
\begin{align*}
\mathcal{L}_{\mathcal{I}_m}&(s) \triangleq \mathbbm{E}\left\{\exp{\left(-s \mathcal{I}_m\right)} \Bigg| \theta, \mathcal{I}_m\right\}\\
&=\mathbbm{E}_{\Phi_u, \left\{|h_{m,i}|^2\right\}}\left\{\exp{\left(-s \sum\limits_{i\in \Phi_u\setminus \{i=0\}}  P_t \left(\frac{c}{4 \pi f d_{m,i}}\right)^2 G |h_{m,i}|^2\right)}\right\}\\
&\stackrel{(a)}=\mathbbm{E}_{\Phi_u}\left\{\prod\limits_{i\in \Phi_u\setminus \{i=0\}} \mathbbm{E}_{|h_m|^2}\left[ \exp{\left(-s  P_t \left(\frac{c}{4 \pi f d_{m,i}}\right)^2 G |h_m|^2\right)}\right] \right\}\\
&\stackrel{(b)}= \mathbbm{E}_{\Phi_u}\left\{\prod\limits_{i\in \Phi_u\setminus \{i=0\}} \left(1+m_2 s  P_t  G\left(\frac{c}{4 \pi f d_{m,i}}\right)^2 \right)^{-m_1}\right\}\\
&\begin{multlined}\stackrel{(c)}=\exp\left(-  \lambda_u \int\limits_0^{2\pi}\int\limits_0^{ \theta_{m, \rm max}}\left[1-\left(1+m_2 s  P_t G \left(\frac{c}{4 \pi f d}\right)^2 \right)^{-m_1}\right] \right.  \left. \vphantom{ \int\limits_0^{2\pi}} \times  R_\oplus^2 \sin{\theta}d\phi d\theta\right),
\end{multlined}
\end{align*}
where step $(a)$ stands since the fading distribution $|h_{m,i}|^2$ for each \ac{IoT} device is \ac{i.i.d.} and is independent of the interfering point process $\Phi_u$. Step $(b)$ follows the \ac{MGF} of the Gamma distribution. Finally, step $(c)$ follows the \ac{PGFL} of the \ac{PPP}, considering that the interference signal originates from devices within the satellite's coverage area.

\section{Proof of Lemma \ref{lemma cov prob}}\label{Proof: lemma cov prob}
Having already derived the expression for availability probability, we now turn our attention to finalizing the successful communication probability by deriving the coverage probability. As previously noted, under the assumption that tier $m$ represents the legitimate tier, it suffices to derive the expression for \(\mathcal{P}_{m, \rm cov}\), the probability that the predefined threshold is lower than the \ac{SINR} of the \ac{LS}. The derivation proceeds are as follows:

\begin{align*}
\mathcal{P}&_{m, \rm cov} \triangleq\mathbb{P}\left[\text{SINR}_{\rm LS}>\beta_{\rm LS}\right]\\
& \stackrel{(a)}=\mathbb{P}\left[\frac{\gamma P_t \left(\frac{c}{4 \pi f d_{m,0}}\right)^2 G |h_{m,0}|^2}{\mathcal{I}_m+\sigma_{\rm S}^2}>\beta_{\rm LS}\right] \\
&=\mathbb{E}_{\theta,\mathcal{I}_m}\left\{1-F_{|h_{m,0}|^2}\left(\frac{\beta_{\rm LS}  (\mathcal{I}_m+\sigma_{\rm S}^2)}{\gamma P_t G}\left(\frac{4 \pi f d_{m,0}}{c}\right)^2\right) \right\}\\
&\begin{multlined}\stackrel{(b)}\approx\mathbb{E}_{\theta,\mathcal{I}_m}\left\{ 1-\left[1-\exp\left(-\frac{(m_1!)^{-\frac{1}{m_1}}}{m_2}\frac{\beta_{\rm LS}}{
    \gamma P_t G}\left(\frac{4 \pi f d_{m,0}}{c}\right)^2 \right. \right. \right. \\ \left.  \left.  \left. \vphantom{\frac{(m_1!)^{-\frac{1}{m_1}}}{m_2}}\times  \left(\mathcal{I}_m+\sigma_{\rm S}^2\right) \right)\right]^{m_1}  \right\}
     \end{multlined}\\
&\stackrel{(c)}= \mathbb{E}_{\theta,\mathcal{I}_m}\left\{ 1- \sum\limits_{q=0}^{m_1}\binom{m_1}{q} (-1)^q \exp{\left(-s_{\rm LS} q \left(\mathcal{I}_m+\sigma_{\rm S}^2\right) \right)} \right\} \\
     &\stackrel{(d)}= \sum\limits_{q=1}^{m_1}\binom{m_1}{q} (-1)^{q+1} \int\limits_0^{\theta_{m, \rm max}}\exp{\left(-q s_{\rm LS}\sigma_{\rm S}^2\right)}\mathcal{L}_{\mathcal{I}_m}\left(q s_{\rm LS}\right)f_{\theta_{m,0}}(\theta)d\theta,
\end{align*}
where step $(a)$ is derived by direct expansion of the definition in (\ref{sinr for LS}), step $(b)$ follows the upper bound for the \ac{CDF} of the Gamma function detailed in (\ref{Upper Gamma}). This approximation is crucial for simplifying the expression involving the channel gain's distribution, facilitating a more tractable analysis of the coverage probability. Step $(c)$ utilizes Newton's generalized binomial theorem to expand the expression. The scaling factor \(s\) in step $(c)$ is defined as:
\begin{align*}
    s_{\rm LS}=\frac{(m_1!)^{-\frac{1}{m_1}}}{m_2}\frac{\beta_{\rm LS}}{
    \gamma P_t G}\left(\frac{4 \pi f d_{m,0}}{c}\right)^2 
\end{align*}
Finally, step $(d)$ integrates the expectation over the distribution of the contact angle, \(\theta_{m,0}\), and the interference model, taking into account the spatial geometry of satellite coverage and the statistical properties of the channel. This concludes the proof.
\section{Proof of Theorem \ref{lemma out prob}}\label{Proof: Lemma Out Prob}
To state the final expression of the secure communication probability, it's sufficient to outline the derivation secrecy outage probability. Here, we consider the influence of potential  \ac{ES}s across all tiers, with the exception of the legitimate tier. The objective is to evaluate the maximum \ac{SINR} experienced by these  \ac{ES}s. Therefore, the secrecy probability of a typical \ac{IoT} device is defined as the probability that the \ac{SINR} by all  \ac{ES}s for all tiers is less than threshold $\beta_{\rm ES}$. First, we will find the strongest \ac{SINR} of any tier $k \neq m$ such that
\begin{align*}
\mathcal{P}_{\rm out}&\triangleq\mathbb{P}\left[\max_{k\neq m} \max_{j\in \Phi_{\rm S_k}}\left\{\text{SINR}_{kj}\right\}<\beta_{\rm ES} \right]\\
&\stackrel{(a)}=\prod_{\substack{k=1 \\ k \neq m}}^{K}    \prod\limits_{j=1}^{N_k} \mathbb{P}\left[\text{SINR}_{kj}<\beta_{\rm ES} \right] \\
&\stackrel{(b)}=  \prod_{\substack{k=1 \\ k \neq m}}^{K} \left[\int\limits_0^{ \theta_{k, \rm max}} \mathbb{P}\left[\text{SINR}_{k}<\beta_{\rm ES} \right] f_{\theta_{k,0}}(\theta)d\theta + \int\limits_{ \theta_{k, \rm max}} ^\pi f_{\theta_{k,0}}(\theta)d\theta \right]^{N_k}\\
\end{align*}
where step $(a)$ follows from the independence of satellites' positions. Step $(b)$ holds because of the \ac{i.i.d.} characteristic of the channel fading.

\par 
Diving into the probability calculation for a single  \ac{ES}, we manipulate the \ac{SINR} condition under the constraint \(d < d_{\text{max}}\), leading to a form that involves expected values over the contact angle distribution and the interference model. The contact angle distribution defined for each satellite can be derived from Lemma \ref{lemma:pdf contact angle} by assuming \(N_k=1\):
\begin{align}
f_{\theta_{k,0}}(\theta)&=\frac{\sin{\theta}}{2}.
\end{align} 
Following similar steps to the coverage probability proof, we arrive at the formulation for the integral part of step $(b)$:
\begin{align*}
    & \int \limits_0^{ \theta_{k, \rm max}} \mathbb{P}\left[\text{SINR}_{S_{k}}<\beta_{\rm ES} \right]f_{\theta_{k,0}}(\theta) d \theta \\  
    & = \sum \limits_{q=0}^{m_1} \binom{m_1}{q} (-1)^q \int\limits_0^{\theta_{k, \rm max}}\exp{\left(-q s_{\rm ES}\sigma_S^2\right)} \mathcal{L}_{\mathcal{I}_{kj}}\left(q s_{\rm ES}\right)f_{\theta_{k,0}}(\theta)d\theta 
\end{align*}
with $s_{\rm ES}$ defined as:
\begin{align*}
    s_{\rm ES}=\frac{(m_1!)^{-\frac{1}{m_1}}}{m_2}\frac{\beta_{\rm ES}}{
    \left[\gamma -\beta_{\rm ES} \left(1-\gamma \right) \right]P_t G}\left(\frac{4 \pi f d_{kj,0}}{c}\right)^2.
\end{align*}
The second integral part in step $(b)$ can be calculated as:
\begin{align*}
    \int\limits_{ \theta_{k, \rm max}} ^\pi f_{\theta_{k,0}}(\theta)d\theta =
    \frac{1+\cos{\theta_{k, \rm max}}}{2}. 
\end{align*}
Combining the above results, we articulate the secrecy outage probability (\ref{eq of out prob}) as
\begin{align*}
       \mathcal{P}_{\rm out} = \prod_{\substack{k=1 \\ k \neq m}}^{K}   & \left[\sum\limits_{q=0}^{m_1}\binom{m_1}{q} (-1)^q  \int\limits_0^{\theta_{k, \rm max}}\exp{\left(-q s_{\rm ES}\sigma_S^2\right)}\mathcal{L}_{\mathcal{I}_{kj}}\left(q s_{\rm ES}\right) \right. \nonumber \\
&\hspace{3cm}  \left. \vphantom{\int\limits_0^{2\pi}} f_{\theta_{k,0}}(\theta)d\theta +\frac{1+\cos{\theta_{k, \rm max}}}{2}\right]^{N_k}.
\end{align*}

\par
In summary, this proof thoroughly derives the secrecy probability by considering the strongest eavesdropper's \ac{SINR} across all non-legitimate tiers, effectively encapsulating the multifaceted dynamics of satellite communication systems where secrecy against  \ac{ES}s is paramount.

\bibliographystyle{IEEEtran}
\bibliography{references}
 
\end{document}